# Sustaining high-fidelity quantum logic in neutral-atom circuits via mid-circuit operations

Rui Lin[1,2,3,*], You Li[1,2,3,*], Le-Tian Zheng[1,2,3,*], Tai-Ran Hu[1,2,3], Si-Yuan Chen[1,2,3], Hong-Ming Wu[1,2,3], Yu-Chen Zhang[1,2,3], Hao-Wen Cheng[1,2,3], Yu-Hao Deng[1,2,3], Zhan Wu[1,2,3], Ming-Cheng Chen[1,2,3], Jun Rui[1,2,3], Chao-Yang Lu[1,2,3], Jian-Wei Pan[1,2,3]

[1]Hefei National Laboratory for Physical Sciences at the Microscale and Department of Modern Physics, University of Science and Technology of China, Hefei 230026, China

[2]Shanghai Branch, CAS Center for Excellence in Quantum Information and Quantum Physics, University of Science and Technology of China, Shanghai 201315, China

[3]Shanghai Research Center for Quantum Sciences, Shanghai 201315, China

Emails: cylu@ustc.edu.cn, pan@ustc.edu.cn

**The realization of fault-tolerant quantum computation[1–4] hinges on the ability to execute deep quantum circuits while maintaining gate fidelities consistently above error-correction thresholds. Although neutral-atom arrays have recently demonstrated high-fidelity two-qubit gates[5–10] and early-stage logical quantum processors[11–16], sustaining such high performance across deep, repetitive circuits remains a formidable challenge due to cumulative motional heating and atom loss. Here we demonstrate a sustainable neutral-atom framework that overcomes these limitations by integrating a suite of hardware-efficient mid-circuit operations. We report a two-qubit controlled logic gate with a raw fidelity of 99.60(1)%, which is further increased to a fidelity of 99.81(1)% via non-destructive erasure detection. Crucially, by implementing in-circuit Raman sideband cooling[17,18] and qubit re-initialization, we demonstrate that gate fidelities can be maintained at the ~99.8% level across multiple operational rounds without observable degradation. By actively managing the internal and motional entropy of the system mid-stream, our in-situ refreshable architecture provides a critical pathway for executing the repeated syndrome-extraction cycles required for large-scale, continuous quantum error correction.**

The roadmap toward fault-tolerant quantum computing[1–4] necessitates scaling across two critical dimensions: the number of qubits and the depth of quantum circuits. In recent years, neutral-atom arrays[19,20] have made remarkable progress by exploiting their reconfigurable connectivity[21] and scalability to demonstrate thousand-atom processors[22–24] and early-stage logical operations[11–16]. As these systems transition from proof-of-principle demonstrations to practical, large-scale algorithms, the second dimension—sustaining high performance over prolonged circuit depths—becomes the primary bottleneck.

In current neutral-atom architectures, the fidelity of entangling gates—typically mediated by Rydberg interactions—is coupled to the motional state of the atoms. During the execution of deep circuits, repetitive gate operations and photon scattering inevitably lead to cumulative motional heating and atom loss. This buildup of entropy acts as a "thermal decay" mechanism: without an active refreshing protocol, gate fidelities decrease rapidly as the circuit progresses. Consequently, even the most advanced demonstrations to date have been largely limited to shallow circuits or a few error-correction cycles[11–16], often requiring the circuit to be terminated and restarted— via entire array reloading and out-of-circuit cooling—to reset the system's entropy[11,12,14,16]. This poses a serious barrier to continuous, long-term quantum operation.

Here, we overcome this "depth barrier" by demonstrating a refreshable neutral-atom framework that sustains high-fidelity quantum logic in deep circuits. As shown in Fig. 1, our approach integrates a hardware-efficient suite of mid-circuit operations— including in-situ Raman sideband cooling, qubit re-initialization, and non-destructive erasure detection—directly into the functional gate sequence. Our experiment begins with loading single $^{87}$Rb atoms into a programmable array of 852-nm optical tweezers, which are subsequently rearranged into a configuration of five well-isolated atom pairs. The assembled atoms are then cooled using D2 Lambda-enhanced gray molasses[25] to

~8 µK in 1-mK-deep traps.

Physical qubits are encoded in the hyperfine clock states $|0\rangle = |F = 1, m_F = 0\rangle$ and $|1\rangle = |F = 2, m_F = 0\rangle$, with long coherence times. We initialize the qubits through a sequence of optical pumping, Raman transfer, and Raman-assisted pumping[5,26], preparing all atoms into the $|0\rangle$ state with fidelity >99%. Coherent single-qubit rotations are driven using Raman laser pulses[27] within the clock-state manifold, achieving fidelity of 99.97%.

Entanglement is mediated through Rydberg interactions. We excite the atoms to the Rydberg state $|70S_{1/2}\rangle$ using a two-photon transition (420+1013 nm), and achieve a Rabi frequency $\Omega/2\pi = 5$ MHz under an intermediate-state detuning of $\Delta/2\pi = 9.1$ GHz (see Figs. 2a and 2b). The two-qubit CZ gate is implemented using a parameterized time-optimal pulse[5,28] (Fig. 2c), in which the 420-nm Rydberg laser is phase-modulated with a sinusoidal profile while its intensity remains constant. We optimize the parameters in the phase profile and extract the gate performance using a global echoed randomized benchmarking sequence (Fig. 2d). This procedure yields a two-qubit gate fidelity of 99.60(1)% (Fig. 2e).

To gain a deeper understanding of the gate's error budget and identify pathways for further optimization, we develop a theoretical model that captures the underlying dynamics of the Rydberg gate operation (see Methods). Our analysis reveals that a significant portion of the infidelity manifests as atom loss, primarily arising from residual Rydberg population due to imperfect gate evolution or black-body radiation induced transitions, which is subsequently expelled by the tweezers. Such findings highlight the critical need for a measurement protocol that can resolve these loss events, enabling physical loss to be converted into erasure errors[29–31] with known locations, which are significantly easier to handle in quantum error correction (QEC). Furthermore, to sustain performance in deep quantum circuits, such a measurement must be non-destructive and mid-circuit compatible[32–34], providing a prerequisite for subsequent refresh operations like in-circuit cooling and re-initialization.

Our loss-resolved measurement employs a two-stage imaging scheme, beginning with a state-selective imaging that picks out the bright state, followed by a state-independent imaging that determines atomic survival. In the first stage, we perform state-selective imaging[35,36] on the closed transition $|F = 2, m_F = 2\rangle \to |F' = 3, m'_F = 3\rangle$, using counter-propagating $\sigma^+$-polarized imaging beams in a finite magnetic field (Fig. 3a). The level splitting and polarization selection provide a clean isolation of this closed transition. Atoms in $|1\rangle$ are rapidly pumped into the stretched state $|F = 2, m_F = 2\rangle$ and continuously scatter photons, whereas atoms in $|0\rangle$ remain dark. As only Doppler cooling is active during this stage, the imaging conditions must be carefully engineered to suppress atom heating and loss. To this end, we strategically strobe the tweezer and imaging beams on and off out of phase, thus avoiding the repulsive excited-state light shift and reducing momentum spread[37]. We further increase the tweezer depth to 2.5 mK to improve the atomic survival.

In the second stage, we resolve the atom presence or loss using a mid-circuit–compatible polarization gradient cooling (PGC) configuration[13] (Fig. 3b). Conventional

PGC methods break down under a finite magnetic field—an essential requirement for mid-circuit operation. To address this, we use one-dimensional counter-propagating $\sigma^+ + \sigma^-$ beams with a relative detuning, which re-establishes effective PGC cooling and imaging. The fluorescence images from both stages are processed using an AI model (Methods). Together, these two stages provide a non-destructive and mid-circuit-compatible measurement that resolves $|0\rangle$, $|1\rangle$, and atom loss, achieving a state-resolved fidelity of 98.9% and an atomic survival probability of 99.7% (Fig. 3c; Methods).

Integrating this loss-resolved measurement mechanism into benchmarking allows us to characterize the intrinsic performance of the entangling gates. By post-selecting experimental trials where both atoms remain trapped after the gate operation, we effectively convert the loss errors into detectable erasure events. This yields a loss-corrected CZ gate fidelity of 99.81(1)% (Fig. 3d), a significant improvement over the raw fidelity of 99.60(1)%. This result not only sets a new benchmark for neutral-atom entangling gates but also highlights the potential for QEC utilizing erasure conversion, where the knowledge of loss locations can significantly relax the fault-tolerant thresholds.

To demonstrate the sustainability of high-fidelity logic in a deep-circuit architecture, we implement a repetitive benchmarking protocol consisting of a cyclic sequence: atom cooling and initialization, gate operations, and non-destructive measurement, after which the sequence returns to the cooling and initialization stage for the subsequent round. As previously noted, the cumulative heating is expected to degrade gate performance over multiple rounds of logic. We empirically validate this by performing repeated randomized benchmarking without active mid-circuit cooling, where the gate fidelity exhibits a drop of ~0.3% from the second round onwards (Fig. 4c). A similar degradation is observed (Fig. 4c) even when we employ local gray molasses cooling (see Methods)—a technique analogous to the PGC imaging in a finite magnetic field—indicating that a more robust entropy removal mechanism is required to refresh the system state.

To address this, we develop a mid-circuit Raman sideband cooling (RSC) protocol that effectively refreshes both the motional and internal states of the qubits. Generally, RSC employs repeated cycles of red-sideband transitions between motional levels, interleaved with dissipative optical pumping, to reduce the mean phonon number, and can in principle cool atoms close to the three-dimensional motional ground state. A distinctive feature of our RSC protocol (Fig. 4a) is that it drives Raman sideband transitions directly between the clock states $|1,0\rangle$ and $|2,0\rangle$—a configuration less commonly adopted in neutral-atom RSC implementations[17,18,38]. This choice makes the cooling process intrinsically robust against inhomogeneities and fluctuations in both the magnetic field and vector light shifts, enabling uniform performance across the atom array. The cooling cycle is closed by Raman-assisted pumping back to $|1,0\rangle$, which simultaneously re-initializes the qubit internal state and provides an integrated mid-circuit "cool-and-reset" operation.

With this mid-circuit RSC protocol in place, we benchmark the CZ gate performance over repeated rounds of circuit execution. The effectiveness of the RSC is evidenced by the consistent sideband spectra throughout the repeated benchmarking process. We achieve a mean phonon number of $\bar{n} \approx 1$ in both the radial and axial directions across all five rounds of operation (Fig. 4b). This sustained cooling performance ensures that the Rydberg gate operates in a regime where Doppler effects and other motional decoherence are effectively mitigated. Consequently, the loss-corrected CZ gate fidelity remains remarkably consistent at ~99.8% across all five consecutive rounds, as shown in Fig. 4c. The transition from a "single-shot" demonstration to a sustained circuit-level operation proves that by effectively refreshing the motional and internal states mid-circuit, the performance of neutral-atom entangling gates can be maintained throughout the execution of a deep quantum circuit.

In summary, we have demonstrated a sustainable architecture for high-fidelity quantum logic in deep neutral-atom circuits by integrating a comprehensive suite of mid-circuit operations. Looking forward, the capability to "refresh" a quantum processor in-situ establishes a critical prerequisite for large-scale fault-tolerant quantum computing. By ensuring that the atoms remain in a near-ground motional state throughout the computation, our architecture provides a robust pathway for executing the repeated syndrome-extraction cycles required by topological codes, such as the surface code.

The entropy-reset strategy demonstrated here can be directly extended to a zoned layout[11,13], where coherent transport[21] and localized addressing naturally enable the independent refreshing of ancilla qubits. Beyond this, our approach is inherently compatible with teleportation-based qubit replacement[39,40], providing an efficient refresh pathway for data qubits by transferring quantum information onto freshly prepared atoms. The sustained, low-entropy operational regime paves the way for the next generation of neutral-atom processors capable of autonomous and continuous quantum error correction.

**References**

1. Shor, P. W. Fault-tolerant quantum computation. in *Proceedings of 37th Conference on Foundations of Computer Science* 56–65 (IEEE Comput. Soc. Press, Burlington, VT, USA, 1996). doi:10.1109/SFCS.1996.548464.

2. Fowler, A. G., Mariantoni, M., Martinis, J. M. & Cleland, A. N. Surface codes: Towards practical large-scale quantum computation. *Phys. Rev. A* **86**, 032324 (2012).

3. Campbell, E. T., Terhal, B. M. & Vuillot, C. Roads towards fault-tolerant universal quantum computation. *Nature* **549**, 172–179 (2017).

4. Preskill, J. Quantum Computing in the NISQ era and beyond. *Quantum* **2**, 79 (2018).

5. Evered, S. J. *et al.* High-fidelity parallel entangling gates on a neutral atom quantum computer. *Nature* **622**, 268–272 (2023).

6. Peper, M. *et al.* Spectroscopy and Modeling of $^{171}$Yb Rydberg States for High-Fidelity Two-Qubit Gates. *Phys. Rev. X* **15**, 011009 (2025).

7. Tsai, R. B.-S., Sun, X., Shaw, A. L., Finkelstein, R. & Endres, M. Benchmarking and Fidelity Response Theory of High-Fidelity Rydberg Entangling Gates. *PRX Quantum* **6**, 010331 (2025).

8. Muniz, J. A. *et al.* High-Fidelity Universal Gates in the $^{171}$Yb Ground-State Nuclear-Spin Qubit. *PRX Quantum* **6**, 020334 (2025).

9. Radnaev, A. G. *et al.* Universal Neutral-Atom Quantum Computer with Individual Optical Addressing and Nondestructive Readout. *PRX Quantum* **6**, 030334 (2025).

10. Senoo, A. *et al.* High-fidelity entanglement and coherent multi-qubit mapping in an atom array. Preprint at https://doi.org/10.48550/arXiv.2506.13632 (2025).

11. Bluvstein, D. *et al.* Logical quantum processor based on reconfigurable atom arrays. *Nature* **626**, 58–65 (2024).

12. Sales Rodriguez, P. *et al.* Experimental demonstration of logical magic state distillation. *Nature* **645**, 620–625 (2025).

13. Bluvstein, D. *et al.* A fault-tolerant neutral-atom architecture for universal quantum computation. *Nature* **649**, 39–46 (2026).

14. Reichardt, B. W. *et al.* Fault-tolerant quantum computation with a neutral atom processor. Preprint at https://doi.org/10.48550/arXiv.2411.11822 (2025).

15. Muniz, J. A. *et al.* Repeated Ancilla Reuse for Logical Computation on a Neutral Atom Quantum Computer. *Phys. Rev. X* **15**, 041040 (2025).

16. Rines, R. *et al.* Demonstration of a Logical Architecture Uniting Motion and In-Place Entanglement: Shor's Algorithm, Constant-Depth CNOT Ladder, and Many-Hypercube Code. Preprint at https://doi.org/10.48550/arXiv.2509.13247 (2025).

17. Kaufman, A. M., Lester, B. J. & Regal, C. A. Cooling a Single Atom in an Optical Tweezer to Its Quantum Ground State. *Phys. Rev. X* **2**, 041014 (2012).

18. Thompson, J. D., Tiecke, T. G., Zibrov, A. S., Vuletić, V. & Lukin, M. D. Coherence

and Raman Sideband Cooling of a Single Atom in an Optical Tweezer. *Phys. Rev. Lett.* **110**, 133001 (2013).

19. Endres, M. *et al.* Atom-by-atom assembly of defect-free one-dimensional cold atom arrays. *Science* **354**, 1024–1027 (2016).

20. Barredo, D., De Léséleuc, S., Lienhard, V., Lahaye, T. & Browaeys, A. An atom-by-atom assembler of defect-free arbitrary two-dimensional atomic arrays. *Science* **354**, 1021–1023 (2016).

21. Bluvstein, D. *et al.* A quantum processor based on coherent transport of entangled atom arrays. *Nature* **604**, 451–456 (2022).

22. Lin, R. *et al.* AI-Enabled Parallel Assembly of Thousands of Defect-Free Neutral Atom Arrays. *Phys. Rev. Lett.* **135**, 060602 (2025).

23. Chiu, N.-C. *et al.* Continuous operation of a coherent 3,000-qubit system. *Nature* **646**, 1075–1080 (2025).

24. Manetsch, H. J. *et al.* A tweezer array with 6,100 highly coherent atomic qubits. *Nature* **647**, 60–67 (2025).

25. Rosi, S. *et al.* Λ-enhanced grey molasses on the $D_2$ transition of Rubidium-87 atoms. *Sci. Rep.* **8**, 1301 (2018).

26. Levine, H. *et al.* Parallel Implementation of High-Fidelity Multiqubit Gates with Neutral Atoms. *Phys. Rev. Lett.* **123**, 170503 (2019).

27. Levine, H. *et al.* Dispersive optical systems for scalable Raman driving of hyperfine qubits. *Phys. Rev. A* **105**, 032618 (2022).

28. Jandura, S. & Pupillo, G. Time-Optimal Two- and Three-Qubit Gates for Rydberg

Atoms. *Quantum* **6**, 712 (2022).

29. Wu, Y., Kolkowitz, S., Puri, S. & Thompson, J. D. Erasure conversion for fault-tolerant quantum computing in alkaline earth Rydberg atom arrays. *Nat. Commun.* **13**, 4657 (2022).

30. Ma, S. *et al.* High-fidelity gates and mid-circuit erasure conversion in an atomic qubit. *Nature* **622**, 279–284 (2023).

31. Scholl, P. *et al.* Erasure conversion in a high-fidelity Rydberg quantum simulator. *Nature* **622**, 273–278 (2023).

32. Deist, E. *et al.* Mid-Circuit Cavity Measurement in a Neutral Atom Array. *Phys. Rev. Lett.* **129**, 203602 (2022).

33. Lis, J. W. *et al.* Midcircuit Operations Using the *omg* Architecture in Neutral Atom Arrays. *Phys. Rev. X* **13**, 041035 (2023).

34. Norcia, M. A. *et al.* Midcircuit Qubit Measurement and Rearrangement in a $^{171}$Yb Atomic Array. *Phys. Rev. X* **13**, 041034 (2023).

35. Martinez-Dorantes, M. *et al.* Fast Nondestructive Parallel Readout of Neutral Atom Registers in Optical Potentials. *Phys. Rev. Lett.* **119**, (2017).

36. Kwon, M., Ebert, M. F., Walker, T. G. & Saffman, M. Parallel Low-Loss Measurement of Multiple Atomic Qubits. *Phys. Rev. Lett.* **119**, (2017).

37. Su, L. *et al.* Fast single atom imaging for optical lattice arrays. *Nat. Commun.* **16**, (2025).

38. Yu, Y. *et al.* Motional-ground-state cooling outside the Lamb-Dicke regime. *Phys. Rev. A* **97**, 063423 (2018).

39. Perrin, H., Jandura, S. & Pupillo, G. Quantum Error Correction resilient against Atom Loss. *Quantum* **9**, 1884 (2025).

40. Baranes, G. *et al.* Leveraging Qubit Loss Detection in Fault-Tolerant Quantum Algorithms. *Phys. Rev. X* **16**, 011002 (2026).

**Figures**

**Fig. 1 | Sustaining high-fidelity quantum logic in deep circuits via mid-circuit operations.** Schematic representation of a refreshable neutral-atom architecture scaling across qubit number (vertical axis) and circuit depth (horizontal axis). As the quantum circuit progresses, repetitive gate operations and photon scattering lead to cumulative motional heating and entropy buildup. This "thermal decay" mechanism rapidly degrades gate performance, visually depicted by the color transition of the physical qubits (circles) and logic gates (rectangles) from cold, high-fidelity states (blue) toward heated, low-fidelity states (red). To address this, we integrate a hardware-efficient suite of mid-circuit operations—including non-destructive measurement (resolving $|0\rangle$, $|1\rangle$, and atom loss), in-situ Raman sideband cooling and qubit re-initialization—to actively remove the accumulated entropy. By effectively refreshing the motional and internal states mid-circuit, the processor recovers to a low-entropy, high-fidelity regime (back to blue), thus enabling continuous, sustainable execution of deep circuits.

**Fig. 2 | Rydberg-mediated entangling gates. a,** Rydberg excitation scheme. Atoms in $|1\rangle$ are coupled to the Rydberg state $|r\rangle = |70S_{1/2}\rangle$ via an intermediate state $|e\rangle =$

$|6P_{3/2}\rangle$, using 420-nm and 1013-nm laser fields with an intermediate-state detuning of $\Delta/2\pi = 9.1$ GHz. **b,** Coherent Rabi oscillations between $|1\rangle$ and $|r\rangle$, exhibiting a Rabi frequency of $\Omega/2\pi = 5$ MHz. **c,** Time-optimal CZ pulse. The 420-nm Rydberg laser is phase-modulated with a sinusoidal profile (orange line), and its intensity remains constant (blue line). **d,** Global echoed randomized benchmarking sequence. Random single-qubit rotations $R_{rand}$ are sampled from a Haar-random distribution, and a final rotation $R_f$ is precomputed to ideally return the population to the initial state $|00\rangle$. A global X pulse is inserted between two CZ gates to echo out single-qubit phases. For each CZ-gate number, a set of 32 single-qubit rotations is applied to decouple single-qubit errors. **e,** CZ gate fidelity. Benchmarking results show the decay of the return probability $P_{|00\rangle}$ as a function of the number of CZ gates. From an exponential fit (asymptote fixed to 0), we extract a CZ gate fidelity of 99.60(1)%. Error bars represent one standard deviation. We also perform a more rigorous randomized benchmarking sequence without echo, which yields a consistent result (see Methods).

**Fig. 3 | Loss-corrected CZ gates via loss-resolved measurement. a,** State-selective closed-transition imaging. Driven by $\sigma^+$-polarized imaging light, atoms in $|1\rangle$ are rapidly pumped into the stretched state $|F = 2, m_F = 2\rangle$ and undergo cycling transitions to $|F' = 3, m'_F = 3\rangle$, continuously scattering photons. Atoms in $|0\rangle$ remain dark. **b,** Mid-circuit PGC imaging. To operate under a finite magnetic field B, counter-propagating $\sigma^+$ and $\sigma^-$ beams are applied with frequency offsets of $\pm\frac{\mu_B g_F B}{\hbar}$ (with $\mu_B$ the Bohr magneton and $g_F$ the Landé factor), thereby re-establishing effective PGC cooling and imaging. **c,** Two-stage imaging correlation for

loss-resolved measurement. We show a scatter plot of the imaging intensities (AI-processed; Methods) from the closed-transition imaging (first stage, horizontal axis) and the mid-circuit PGC imaging (second stage, vertical axis), with marginal distributions shown on the side panels. Data are taken with atoms prepared in $|F=2\rangle$ (orange) and $|F=1\rangle$ (grey). Red dashed lines indicate the classification thresholds. The data clusters clearly distinguish between the qubit states $|1\rangle$ (right), $|0\rangle$ (upper left), and atom loss (lower left), with event counts labeled for each region. **d,** Loss-corrected CZ gate fidelity. By post-selecting trials where both atoms remain trapped, we extract a loss-corrected CZ gate fidelity of 99.81(1)% (exponential fit; asymptote fixed to 1/4). Error bars represent one standard deviation.

**Fig. 4 | Sustaining CZ gates across multiple mid-circuit rounds. a,** Raman sideband cooling (RSC) configuration. Left: Geometry and polarizations of the three RSC beams (RSC1–3) relative to the atom array and the magnetic field. RSC1 and RSC3 address the radial motional mode; RSC1 and RSC2 address both the radial and axial motional modes. Right: Level diagram for the RSC protocol. Raman sideband transitions are driven directly between the clock states $|1,0\rangle$ and $|2,0\rangle$, with each cooling cycle removing one phonon ($|n\rangle \to |n-1\rangle$). The cycle is closed by Raman-assisted pumping (RAP) back to $|1,0\rangle$. **b,** Sustained RSC performance. A representative Raman sideband spectrum is shown, with five clearly resolved features: the carrier (black) is centered at zero detuning, while the inner (outer) sidebands correspond to the axial (radial) motional mode. The pronounced asymmetry between the red and blue sidebands for both modes indicates efficient cooling. The inset shows the extracted mean phonon

number $\bar{n}$ for the axial (green) and radial (orange) modes, maintained at $\bar{n} \approx 1$ across all five mid-circuit rounds. **c,** Sustainability of CZ gate fidelity. Loss-corrected CZ gate fidelities are measured over five consecutive mid-circuit rounds under three distinct conditions. With mid-circuit RSC (blue), the gate fidelity is sustained at a consistent level of ~99.8% without observable degradation. In contrast, the gate fidelity with local gray molasses (Local GM, orange) or without active cooling (green) exhibits significant drops from the second round onwards. Error bars represent the standard error of the exponential fit.

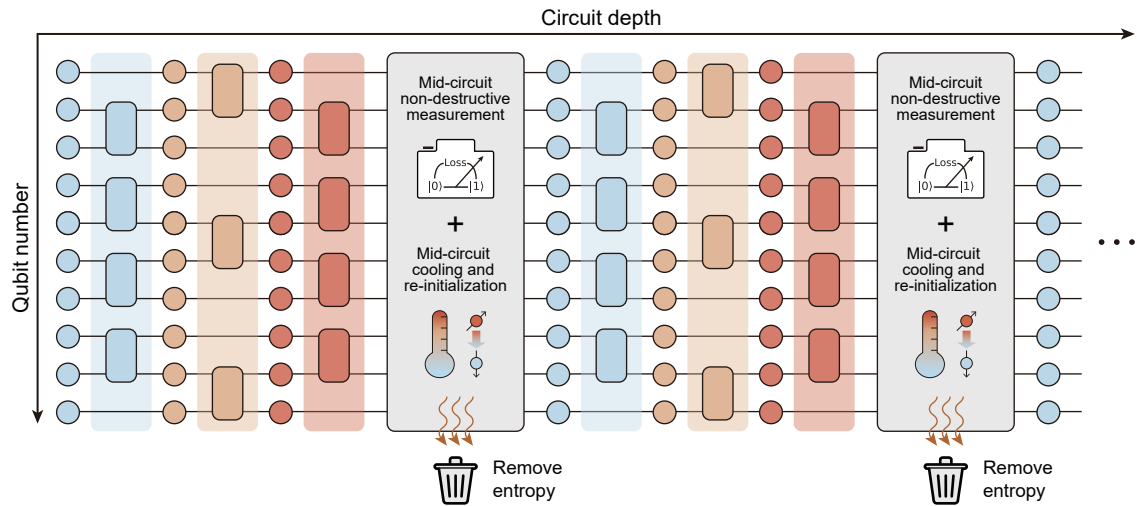

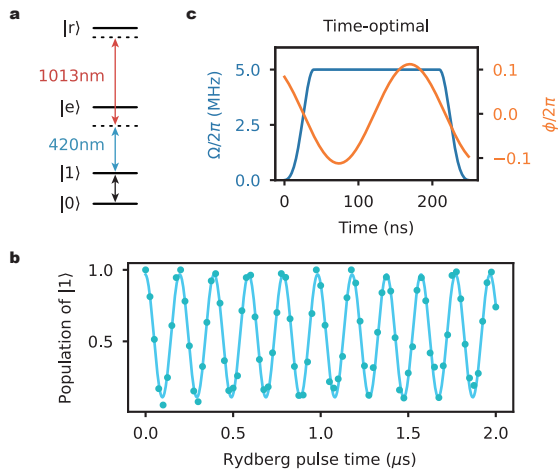
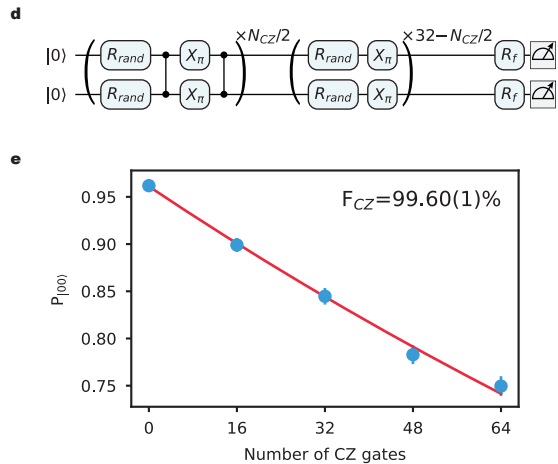

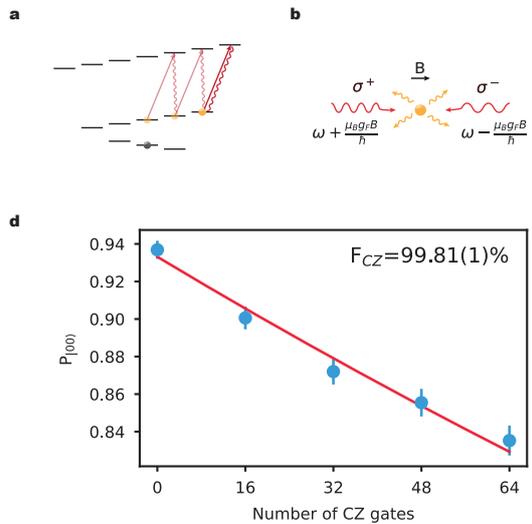
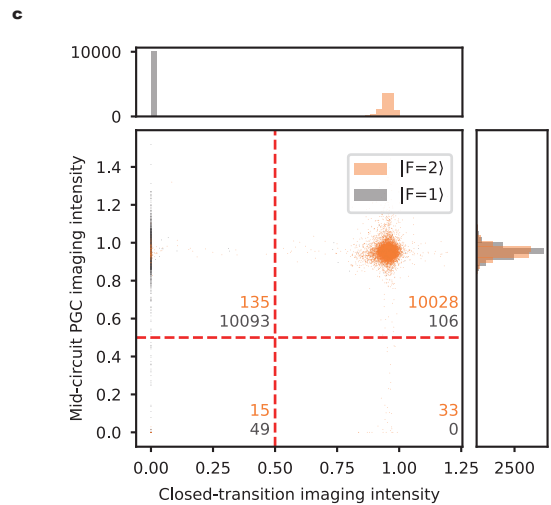

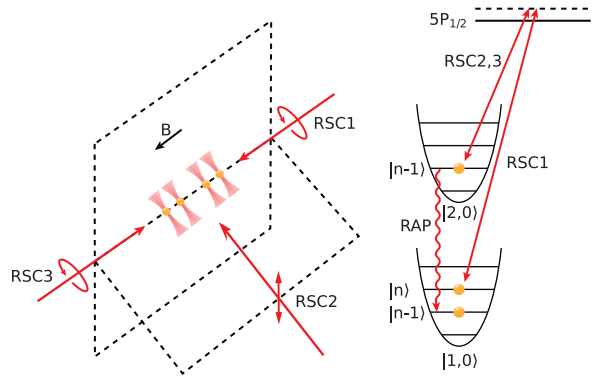
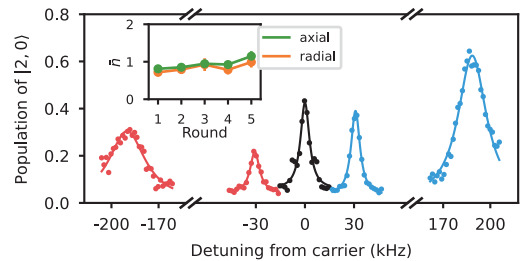
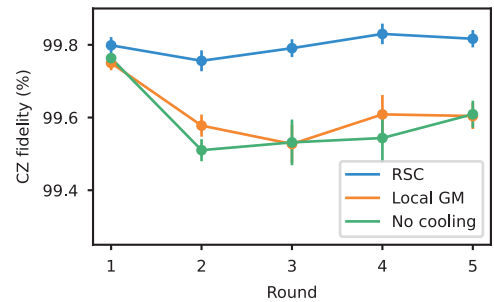